\documentclass{PoS}
\usepackage[FIGTOPCAP,nooneline]{subfigure}
\usepackage{mathrsfs}
\usepackage{amsmath}
\usepackage{caption}
\makeatletter
\newcommand{\figcaption}[1]{\def\@captype{figure}\caption{#1}}
\newcommand{\tblcaption}[1]{\def\@captype{table}\caption{#1}}
\makeatother
\usepackage{wrapfig}
\usepackage{cite}
\usepackage{here}

\title{$K_{l3}$ form factors in $N_f = 2+1$ QCD at physical point on large volume}

\ShortTitle{$K_{l3}$ form factors in $N_f = 2+1$ QCD at physical point on large volume}

\author{ 
{\speaker{J. Kakazu}$^{1}$, K.-I. Ishikawa$^{2,3}$, N. Ishizuka$^{1,4}$, Y. Kuramashi$^{1,4}$,
Y. Nakamura$^{5}$, \  \  \  \  \  \  \  \  \  \  \  \  \  \ Y. Namekawa$^{6}$, Y. Taniguchi$^{1,4}$, N. Ukita$^{4}$,
T. Yamazaki$^{1,4}$, and T. Yoshie$^{1,4}$}\\%

\vspace{-5mm}
 \begin{center}
 {\normalsize \bf \sffamily (PACS Collaboration)}
 \end{center}
        $^1$Graduate School of Pure and Applied Sciences, University of Tsukuba, Tsukuba, Ibaraki 305-8571, Japan\\
        $^2$Graduate School of Science, Hiroshima University, Higashi-Hiroshima, Hiroshima 739-8526, Japan\\
        $^3$Core of Research for the Energetic Universe, Hiroshima University, Higashi-Hiroshima 739-8526, Japan\\
        $^4$Center for Computational Sciences, University of Tsukuba, Tsukuba, Ibaraki 305-8577, Japan\\
        $^5$RIKEN Center for Computational Science, Kobe, Hyogo 650-0047, Japan\\       
        $^6$Institute of Particle and Nuclear Studies,
High Energy Accelerator Research Organization (KEK), Tsukuba 305-0801, Japan  \\
        \\
        E-mail: \email{kakazu@het.ph.tsukuba.ac.jp}}

\abstract{We present our results of the $K_{l3}$ form factors on the volume whose spatial extent is more than $L=$10 fm,
 with the physical pion and kaon masses using the stout-smearing clover $N_f = 2+1$ quark action 
 and Iwasaki gauge action at $a^{-1}\approx2.3$ GeV. The $K_{l3}$ form factor at zero momentum transfer 
 is obtained from fit based on the next-to-leading (NLO) formula in SU(3) chiral perturbation theory. 
 We estimate systematic errors of the form factor, mainly coming from the finite lattice spacing effect.
 We also determine the value of $|V_{us}|$ by combining our result with the experiment  and check the consistency with the standard model prediction. 
 The result is compared with the previous lattice calculations.}

\FullConference{37th International Symposium on Lattice Field Theory - Lattice2019\\
		16-22 June 2019\\
		Wuhan, China}

\begin{document}

\section{Introduction}
Semileptonic decay of kaon into pion and lepton-neutrino pair, so-called $K_{l3}$ 
decay plays an important role to determine $V_{us}$, 
which is one of the Cabbibo-Kobayashi-Maskawa (CKM) matrix~\cite{CKM} elements
to explain mixture between up and strange quarks. 
$ |V_{us}|$ is necessary to examine the existence of
physics beyond the standard model (BSM).
In the standard model, the unitary condition
of the up quark part in the CKM matrix leads
to vanishment of $\Delta_u \equiv |V_{ud}|^2+|V_{us}|^2+|V_{ub}|^2 - 1$. 
Since $|V_{ud}|$ has been obtained accurately
and $|V_{ub}|$ is tiny, high precision
determination of $|V_{us}|$ is required for a check of
 $\Delta_u = 0$ or not.

At present, the most precise determination of $|V_{us}|$  
is the combination of experimental results and lattice calculations~\cite{PDG,FLAG}.
However, there are some uncertainties, 
for instance, chiral extrapolations to the physical pion
and kaon masses and finite size effect. 
To recognize the BSM signature,  
these uncertainties should be reduced.
Hence we determine the $K_{l3}$ form factor
in dynamical $N_f=2+1$ lattice QCD calculation with the physical quark masses
on the spacial volume of more than (10fm)$^3$.
The detailed analyses in this study are presented in Ref.~\cite{Vus}.

\section{Calculation of form factors}
The $K_{l3}$ form factors $f_+(q^2)$ and $f_-(q^2)$ are defined by the matrix element of the weak vector current as,
\begin{eqnarray}
\label{def:ff}
\langle K (\vec{p}') \left | V_{\mu}  \right | \pi(\vec{p}) \rangle = ({p'}+{p})_{\mu}f_{+}(q^2)+ ({p'}-{p})_{\mu}f_{-}(q^2),
\end{eqnarray} 
where $V_{\mu}$ is weak vector current and $q=p'-p$ is the momentum 
transfer.
The scalar form factor $f_0(q^2)$ is defined by combination of vector form factors,
\begin{eqnarray}
f_{0}(q^2) =f_{+}(q^2) + \frac{-q^2}{({m^2_{K}}-{m^2_{\pi}})}f_{-}(q^2)
= f_{+}(q^2)\left(1+ \frac{-q^2}{({m^2_{K}}-{m^2_{\pi}})}\xi(q^2)\right), 
\label{eq:f0}
\end{eqnarray}
where $\xi(q^2)=f_{-}(q^2)/f_{+}(q^2)$.
At $q^2=0$, the two form factors, $f_+(q^2)$ and $f_0(q^2)$ 
give the same value, $f_+(0)=f_0(0)$.

To obtain the form factors, we calculate the meson
3-point function with the weak vector current $C_{\mu}^{\pi K}(\vec{p}, \vec{p}', t)$, which is given by
\begin{eqnarray}
C_{\mu}^{\pi K}(\vec{p}, \vec{p}', t)&=&
\langle 0 | O_K( \vec{p}',t_f) V_\mu(\vec{q},t) O^\dagger_{\pi}( \vec{p},t_i)| 0 \rangle
\\
&=&\frac{Z_{K}(\vec{p}')Z_{\pi}(\vec{p})}{4E_{K}(\vec{p}')E_{\pi}(\vec{p})}\frac{1}{Z_V}
\langle K (\vec{p}') \left | V_{\mu} \right | \pi(\vec{p}) \rangle   e^{{-E_{K}(\vec{p}')(t_f-t)} }e^{{-E_{\pi}(\vec{p})(t-t_i)}}+\cdots,
\end{eqnarray}
where $Z_V$ is the renormalization factor of the vector current, 
and $t_i < t < t_f$. $E_{\pi}$ and $E_{K}$ denote the energy of pion and kaon, respectively.
Their energies are determined by the equation $E_X= \sqrt{m_X^2 +({ \frac{2\pi}{L}\vec{ n}})^2}$ 
using the fitted mass $m_X$ with the label $X$ assigned to $\pi$ or $K$,
where $L$ is the spatial extent and $\vec{n}$ is integer vector 
which represents the direction of meson's momentum.
$m_{X}$ and $Z_{X}(\vec{0})$ are evaluated from 
the meson 2-point functions given by
\begin{eqnarray}
C_{}^{X}( \vec{0},t)=\langle 0 |O_X( \vec{0},t)O^{\dagger}_X( \vec{0},t_i) |0\rangle= \frac{|Z_X( \vec{0} )|^2 }{2m_X} 
( e^{-m_X |t-t_i|}+e^{-m_X (2T-|t-t_i|)})+\cdots,
\end{eqnarray}
where $T$ is the temporal extent. The periodicity in the temporal direction is effectively doubled
thanks to averaging the 2-point functions with the temporal periodic and anti-periodic conditions.
The terms of the dots ($\cdots$) denote the contributions from excited states.

For construction of the form factors,  
we define the quantity $R_\mu(\vec{p}, \vec{p}', t; t_{sep})$ which consists of 2- and 3-point functions 
in $t_{sep}\equiv|t_f-t_i|$,
\begin{eqnarray}
\label{eq:ratio1}
R_\mu(\vec{p}, \vec{p}',t ; t_{sep})=\frac{C_{\mu}^{ \pi K}(\vec{p}, \vec{p}', t)}{C_{}^{\pi}( \vec{p},t)C_{}^{K}( \vec{p}',t)}\ 
\xrightarrow{t_i \ll t \ll t_f} \  A_{\mu}(\vec{p},\vec{p}')+B_{\pi,\mu}(\vec{p},t)+B_{K,\mu}(\vec{p}',t)+\cdots.
\end{eqnarray} 
The term $ A_{\mu}$ is proportional to matrix element $\langle K (\vec{p}') \left | V_{\mu} \right | \pi(\vec{p}) \rangle$, 
and the terms $B_{X,\mu}$ are the first excited state contributions of meson $X$ 
and ($\cdots$) denotes the other excited state contributions.
After extracting $ A_{\mu}$, we could obtain the vector form factor 
$f_+(q^2)$ and $f_-(q^2)$ by the combination with the relation Eq.~(\ref{def:ff}).

\section{Simulation setup}
We use the configurations which were generated at the physical point, 
$m_\pi =0.135$ GeV, on the large volume corresponding to
$La=Ta=10.9$ fm ($L=T=128$), which is a part of 
the PACS10 configuration~\cite{PACS10}.
The configurations were generated by using $N_f=2+1$ 
non-perturbative Wilson clover quark action with six stout smearing link~\cite{ref.conf1} 
(smearing parameter $\rho=0.1$) and the improvement coefficient
 $c_{SW}=1.11$, and the Iwasaki gauge action~\cite{ref.conf2} at $\beta =1.82$
corresponding to $a^{-1}=2.3162(44)$ GeV.
The hopping parameters of two degenerate light quarks and strange quark are 
$(\kappa_{ud},\kappa_{s})= (0.126117, 0.124902)$, respectively.

In the calculation of the form factors, we use 20 configurations in total. 
We adopt 8 sources in time with 16 time separation
per configuration, and 4 choices of the temporal axis thanks
to the hypercube lattice. In the calculation of 2- or 3-point functions, we use 
$\mathbb{Z}(2) \otimes  \mathbb{Z}(2)$ random wall source which is spread in
the spatial sites, and also color and spin spaces~\cite{ffL3}.
We choose the three temporal separations $t_{sep}= 36, 42, 48$ ($\approx 3.05, 3.55, 4.05$ fm)
to dominate $ A_{\mu}(\vec{p},\vec{p}')$ for sufficiently large $t_{sep}$.
One random source is used in the calculations of $t_{sep}=36$
and two sources for the others. 
The 3-point function is calculated using the sequential source technique
at the sink time slice $t_f$, where the meson momentum is fixed to zero.
We calculate $C_{\mu}^{\pi K}(\vec{p}, \vec{p}', t)$ and $C{}^{X}(\vec{p},t)$
with the momentum $\vec{p}=(2\pi/L)\vec{n}$ of
$n\equiv|\vec{n}|^2 \le 6$.
For suppression of the wrapping around effect 
of the 3-point functions, which is similar to that in Ref.~\cite{ref.wrap}, we average the 3-point functions with the periodic 
and anti-periodic boundary conditions in the temporal direction.

\section{Result}
We extract $A_{\mu}$ by fitting $R_\mu$ at each momentum transfer, 
$q_n^2=-(m_K^2+m_{\pi}^2-2m_K \sqrt{m_{\pi}^2 +n( \frac{2\pi}{L})^2} \ )$,
with the fit form considering first excited state contributions
\begin{eqnarray}
\label{eq:exfit}
R_{\mu}(\vec{p}, \vec{p}',t ; t_{sep})=A_{\mu}+B_{\mu}\exp(-t({\Delta E_\pi}))+C_{\mu}\exp(-(t_f-t)({\Delta m_K})),\\
\Delta E_\pi=(E_{\pi*}-E_{\pi}),\  \Delta m_K=(m_{K*}-m_{K}),\nonumber
\end{eqnarray}
where $E_{\pi*}= \sqrt{m_{\pi*}^2 +({\frac{2\pi}{L}\vec{n}})^2} $ is the energy of the first excited state pion.
We apply the experimental values of first excited state pion and kaon 
$m_{\pi*}=1.30$GeV and $m_{K*}=1.46$GeV, respectively~\cite{PDG}.

\begin{figure}[H]
 \begin{center}
\subfigure[Temporal component $A_{4}$]{\includegraphics*[width=72mm]{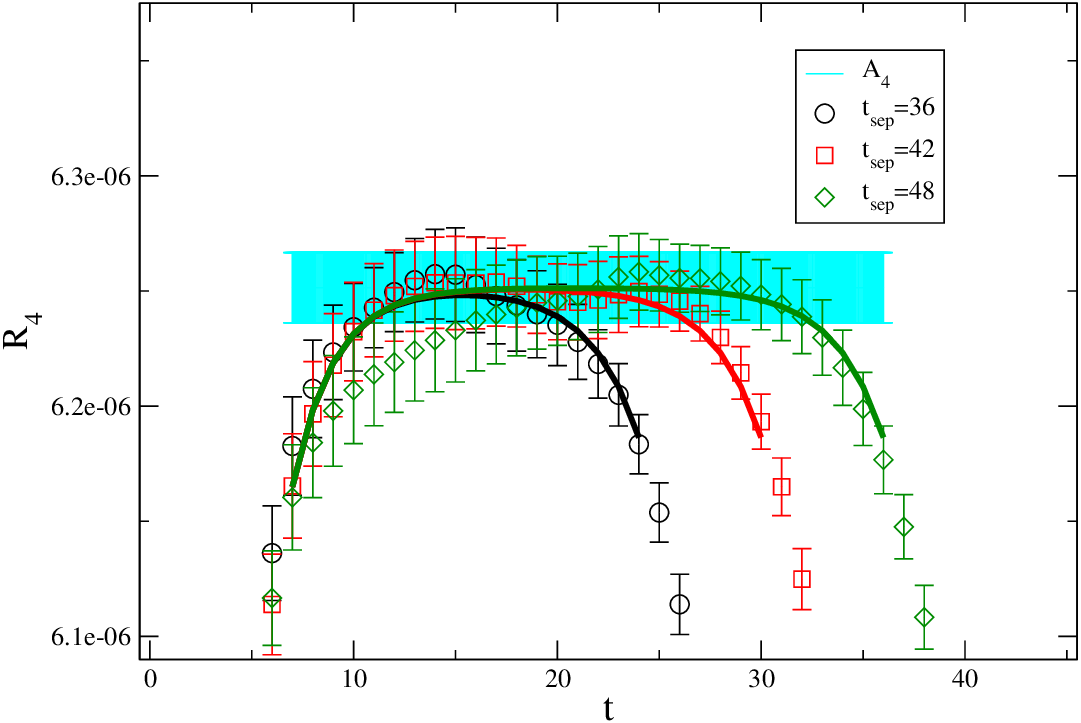}
\label{fig:A4}}
\subfigure[Spatial component $A_{i}$]{\includegraphics*[width=72mm]{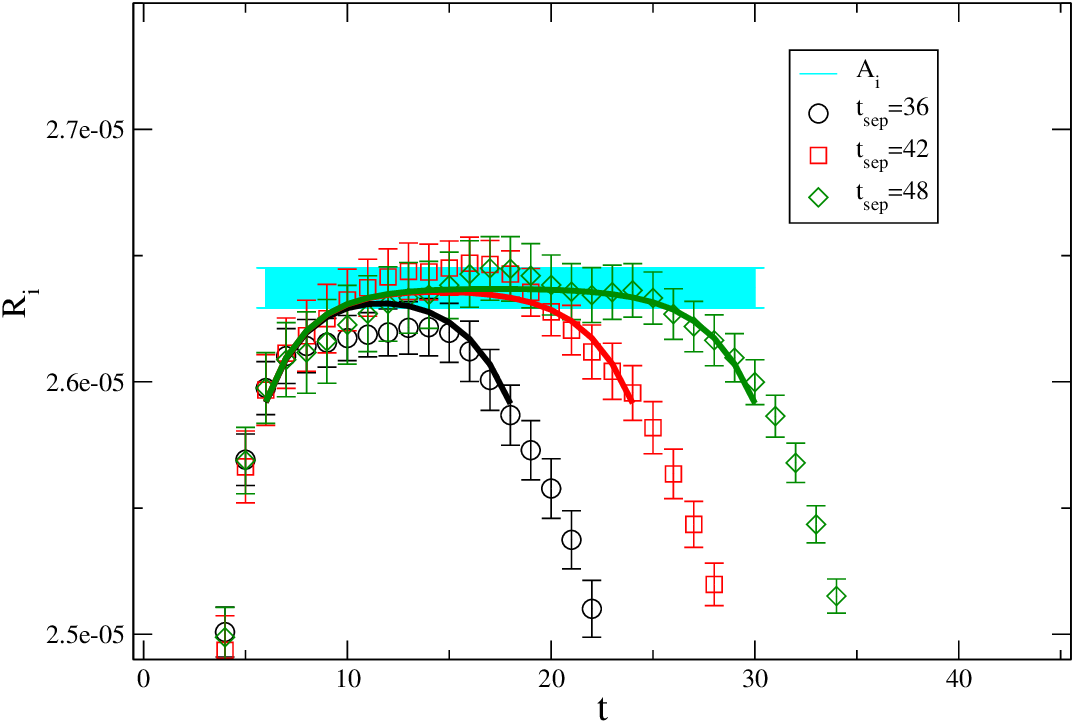}
\label{fig:Ai}}
  \end{center}
 \caption{Extraction of $A_\mu$ from the quantity $R_\mu$ when $|\vec{p}| = 2\pi/L$. 
 Black circle, red square and green diamond symbols in both 
 panels represent the quantities {$R_\mu(\vec{p}, \vec{0}, t ; 36)$}, 
 $R_\mu(\vec{p}, \vec{0}, t ; 42)$, $R_\mu(\vec{p}, \vec{0}, t ; 48)$, respectively. Bold fit curves are drawn by Eq.~(\ref{eq:exfit}).
 Cyan bands represent $A_\mu$ with the statistical error.
 }
\label{fig:fitex}
  \end{figure}

Figure~\ref{fig:fitex} indicates the fit result at $\vec{p} = 2\pi/L$ 
($q_1^2=-(m_K^2+m_{\pi}^2-2m_K \sqrt{m_{\pi}^2 +(\frac{2\pi}{L})^2} \ )$) 
combined all the $t_{sep}$ with the fit form in Eq.~(\ref{eq:exfit}).
We use the fit range of  $7\le t \le(t_{sep}-12)$ in $R_4$ 
(the left panel in Fig.\ref{fig:fitex}),
and use the range of  $6\le t \le(t_{sep}-18)$ in $R_i$ (the right panel in Fig.\ref{fig:fitex}).

After constructing the form factors from the extracted $A_{\mu}$, 
we employ formulae based on the NLO SU(3) ChPT~\cite{ChPT} for interpolation to $q^2=0$
\begin{eqnarray}
\label{eq:chf+}
f_+(q^2)&=&1-2\frac{q^2}{F_0^2}L_9(\mu)+\frac{3}{2}H_{K\pi}(-q^2)+\frac{3}{2}H_{K\eta}(-q^2)+c^{+}_0+c^{+}_2(-q^2)^2, \\
\label{eq:chf0}
f_0(q^2)&=&1-4\frac{q^2}{F_0^2}L_5(\mu)
-\frac{1}{8\pi^2 F_0^2}\left(5 q^2+ 2 \Sigma_{\pi K} - \frac{3 \Delta_{\pi K}^2}{q^2}\right)
\bar{J}_{\pi K}(-q^2) \nonumber\\
&-& \frac{1}{24\pi^2 F_0^2}\left( 3 q^2 + 2 \Sigma_{\pi K} - \frac{\Delta_{\pi K}^2}{q^2}\right)
\bar{J}_{K \eta}(-q^2)- \frac{q^2}{4 \Delta_{\pi K}} \left( 5 \mu_\pi- 2 \mu_K - 3 \mu_\eta \right) \nonumber\\
 &+&c^{0}_0+c^{0}_2(-q^2)^2,
\end{eqnarray}
where $\Sigma_{XY} = m_X^2 + m_Y^2$, $\Delta_{XY} = m_X^2 - m_Y^2$,
$\mu_X=\frac{m_X^2}{32\pi^2 F_0^2}ln\left(\frac{m_X^2}{\mu^2}\right)$, and 
$L_9(\mu),\  L_5(\mu)$ are the {low energy constants of the NLO SU(3) ChPT}.
$H_{XY}$, $\bar{J}_{XY}$, $(X,Y=\pi, K, \eta)$ represent one-loop integrals in Ref.~\cite{ChPT}. 
$\mu=0.77$ GeV is the renormalization scale. 
$F_0=0.079$ GeV is the decay constant in the SU(3) chiral limit.
The NLO ChPT formulae have only trivial terms at $q^2=0$, 
thus we add the nontrivial term $c^{+,0}_0$. We also add the 
$c^{+,0}_2(-q^2)^2$ as the NNLO analytic term has $-q^2$ dependence.
\begin{figure}[H]
 \begin{center}
\subfigure[$q^2$ dependence]{\includegraphics*[width=72mm]{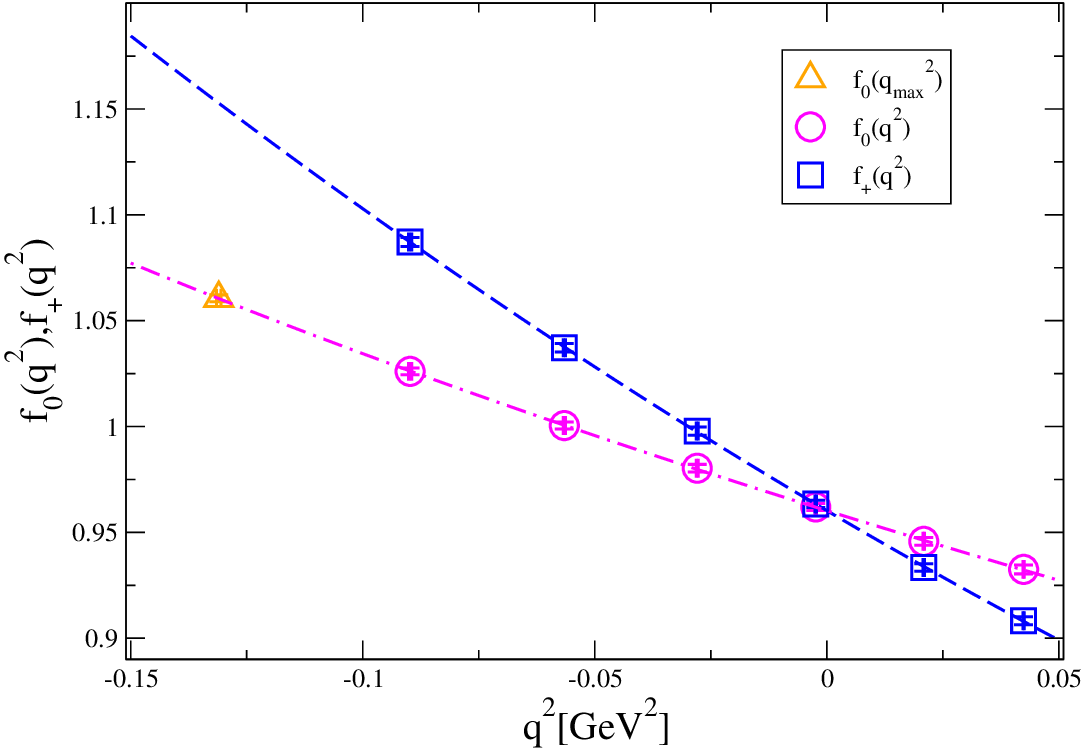}
      \label{fig:ffq16}}
\subfigure[near $q^2=0$ region]{\includegraphics*[width=72mm]{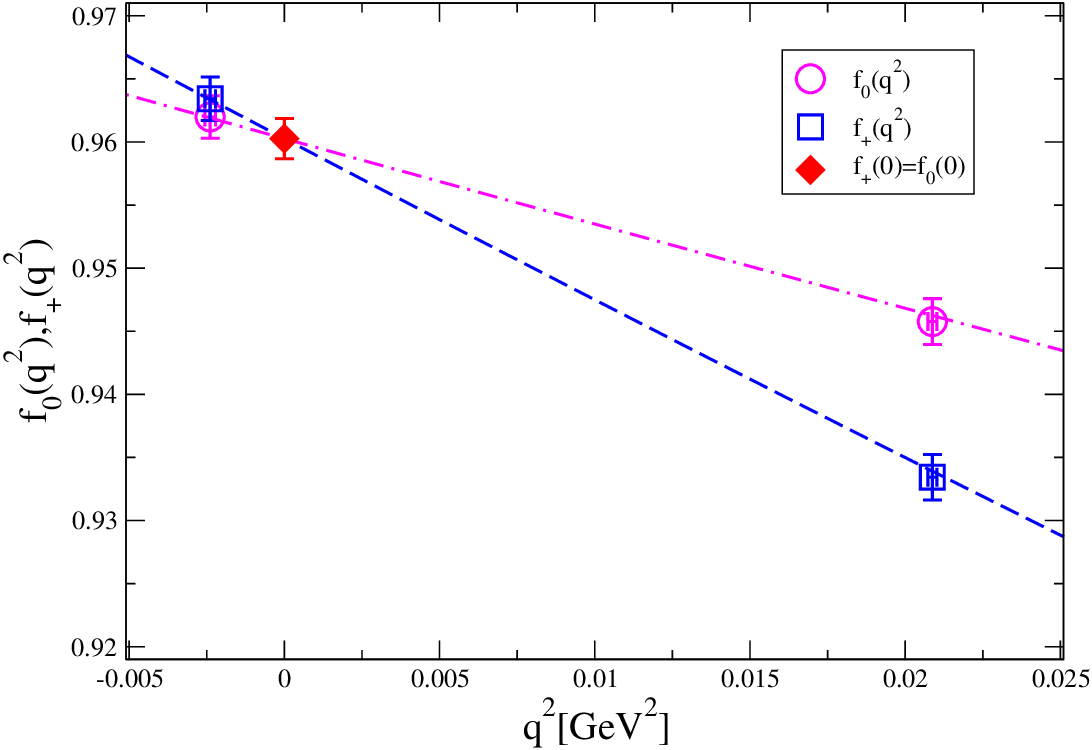}
  \label{fig:q0}}
  \end{center}
 \caption{(a) $q^2$ dependence of $K_{l3}$ form factors.
The square and circle symbols denote $f_+(q^2)$ and $f_0(q^2)$, respectively.
The dashed and dot-dashed curves represent the simultaneous fit result of
$f_+(q^2)$ and $f_0(q^2)$, respectively, with the ChPT forms in
Eqs.(\protect\ref{eq:chf+}) and (\protect\ref{eq:chf0}) added the analytic terms.
The orange triangle at the far left in Fig. (a) represents $f_0(q_{min}^2)$, 
where $q_{min}^2=-(m_K-m_{\pi})^2$.
The diamond symbol denotes the fit result of $f_+(0)=f_0(0)$.
Fig. (b) is the magnification of the left panel near $q^2=0$
\label{fig:fitq}  
}
  \end{figure}

Figure~\ref{fig:fitq} shows the simultaneous fit result of
$f_+(q^2)$ and $f_0(q^2)$ using the ChPT forms in Eqs.~(\ref{eq:chf+}) and~(\ref{eq:chf0}).
The datum of $f_0(q^2_{min})$ 
is not included in the fit.
We also check that the inclusion of the datum 
in the fit does not change the fit result qualitatively. 
The ChPT forms well describe our data, and
$\chi^2/{\rm d.o.f.}\approx 0.06$ in the uncorrelated fit.

We investigate systematic errors in $f_{+,0}(0)$.
The largest error is from the discretization.
The discrepancy of $Z_V$ is regarded as a systematic error
coming from the finite lattice spacing effect, and it must vanish in the continuum limit.
We obtain $Z_V$ of weak current using two methods.
One is determination of $Z_V$ using electromagnetic current conservation. 
We obtain the quantities from the combination of 3-point functions of $t_{sep}=36$ with 
electromagnetic current and 2-point functions of pion and kaon, 
and estimate by geometric mean of these.
This yields $Z_V=0.95586(18)$.
The other is Schr\"{o}dinger functional method, resulting in $Z_V = 0.95153(76)$~\cite{ZV}. 
The difference of the form factor by choice of $Z_V$ is 0.45\%.
The systematic error from finite size effects could be 
ignored because of $O(e^{-m_{\pi}L})\approx 0.06 \%$.

Our result of the form factors with systematic error is 
\begin{eqnarray}
f_{+,0}(0)=0.9603(16)(44),
\end{eqnarray}
where the first error is statistical error and the second is the discrepancy of $Z_V$. 

\begin{figure}[H]
\centering
\includegraphics*[width=115mm]{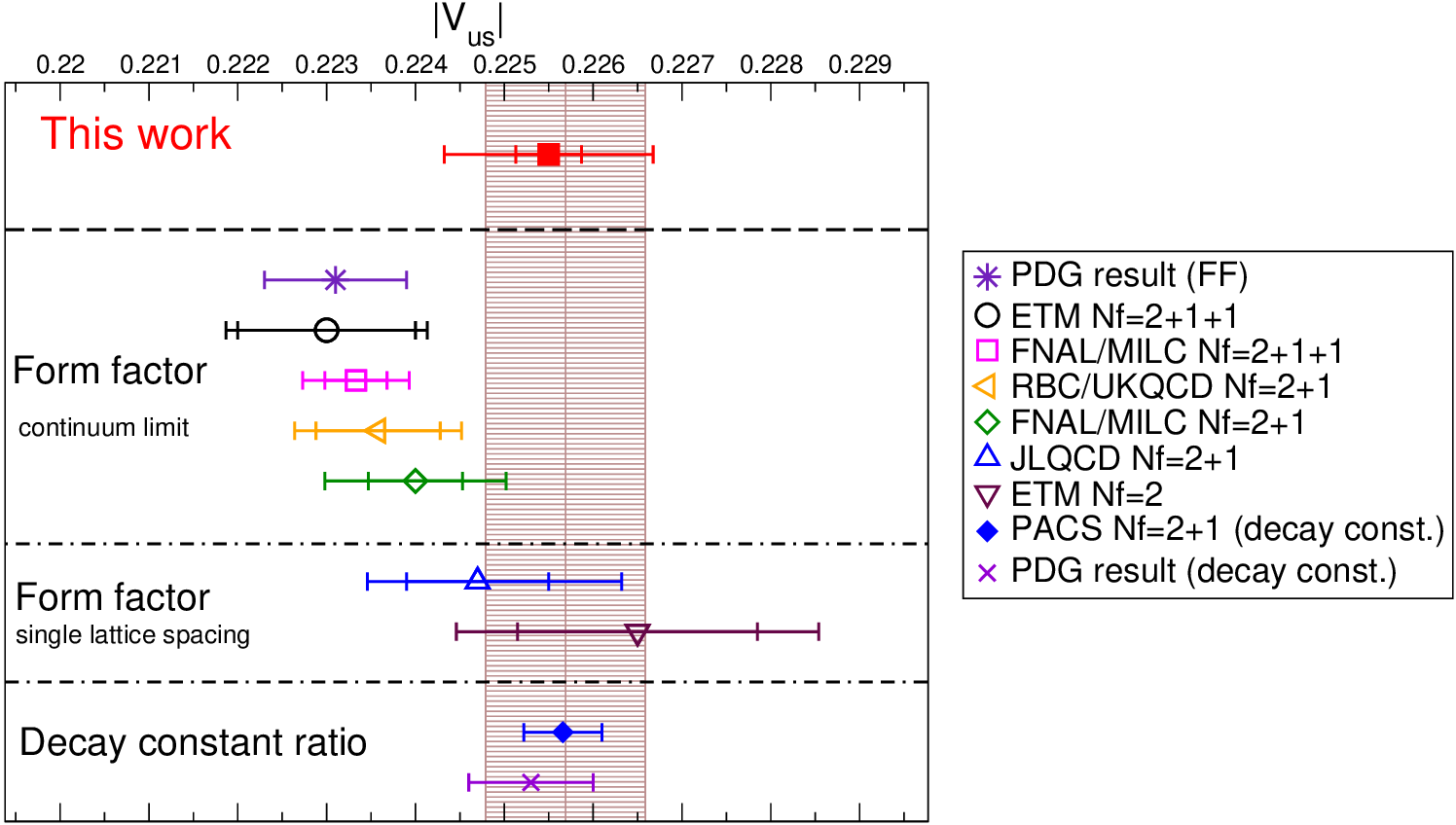}
\caption{Comparison of $|V_{us}|$. 
The filled square symbol is our result 
from the $K_{l3}$ form factor with the ChPT fit.
The filled diamond symbol is estimated by the decay constant ratio 
calculated with the same configuration~\cite{PACS10}.
The star and cross symbols express the PDG's results~\cite{PDG}
from the $K_{l3}$ form factor and the ratio of the decay constant, respectively.
Brown band is the SM prediction from $\Delta_u=0$ and from Ref.~\cite{Vud}. 
The other open symbols represent previous lattice QCD results from
the $K_{l3}$ form factor~\cite{ffL1,ffL2,ffL3,ffL4,ffL5,ffL6}. 
  \label{fig:Vus}
}
 \end{figure}

After interpolating to $q^2=0$ we estimate
the absolute value of the CKM matrix element $|V_{us}|$ 
by combining the value $|V_{us}|f_+(0)=0.21654(41)$ 
derived from the $K_{l3}$ decay rate~\cite{ffexp} as
 \begin{eqnarray}
   |V_{us}|= 0.22550(37)(103)(43),
\end{eqnarray}
where the first error is statistical error, and the second comes
from the choice of $Z_V$ and the third comes from experiment. 
Figure~\ref{fig:Vus} shows the comparison of our results
from the ChPT with the PDG's estimations~\cite{PDG},
other lattice results~\cite{ffL1,ffL2,ffL3,ffL4,ffL5,ffL6},
and the one from the unitarity condition $\Delta_u = 0$.
Our result with total error covers the SM prediction which is estimated  
by the combination $\Delta_u = 0$ and $|V_{ud}|$ from Ref.~\cite{Vud}
(ignoring $|V_{ub}|$ because of too small effect). 
The largest deviation from the previous lattice calculation of continuum limit is $1.8\sigma$
\footnote{The result of the latest determination of $|V_{ud}|=0.97370(14)$~\cite{Vud2},
 the $|V_{us}|$ estimation from $\Delta_u = 0$ combining this value is 
$|V_{us}|=0.22783(59)$. The difference from our result is 1.7 $\sigma$.}
.

We also confirm the consistency on the slope of the form 
factors between our results and  experimental results.
We estimate the slopes of the form factors $\lambda_{+,0}$ defined by
\begin{eqnarray}
\lambda_{s}=\left. \frac{m^2_{\pi_\pm{\rm phys}}}{f_{s}(0)}\frac{df_{s}(t)}{dt} \right|_{t=-q^2=0},
\end{eqnarray}
where label $s$ is assigned to ${+}$ or ${0}$.
Our results from the ChPT fit are
\begin{eqnarray}
\lambda_{+}=2.62(4)\times 10^{-2}, & \lambda_{0}=1.38(4)\times 10^{-2}.& 
\end{eqnarray}
They are consistent with the experimental results 
$\lambda_{+}=2.58(4)\times 10^{-2}$ and $\lambda_{0}=1.36(7)\times 10^{-2}$ ~\cite{ffexp}.

\section{Summary}
We present the results of 
the $K_{l3}$ form factors $f_+(q^2)$ and $f_0(q^2)$
in $N_f=2+1$ lattice QCD at the physical point 
($m_{\pi} \approx 0.135$ GeV) on 
the large volume lattice with (10 fm)$^4$.
Since we calculate the form factors in close to zero momentum transfer,
we perform an interpolation to $q^2=0$ by using
the NLO SU(3) ChPT with NNLO analytic terms.
Using our result of the form factor at $q^2=0$, 
$|V_{us}|$ is estimated by combining the form factor with the $K_{l3}$ decay rate.
Our result is consistent with the SM prediction and is slightly 
larger than previous lattice results in the continuum limit. 
The choice of $Z_V$, which is regarded as the systematic
error from the finite lattice spacing effect,
is the largest systematic error. 
Thus, 
an important future work 
is to evaluate form factors at one or more
finer lattice spacings for taking the continuum limit.

\section*{Acknowledgement}
Numerical calculations in this work were performed on the Oakforest-PACS 
at Joint Center for Advanced High Performance Computing under Multidisciplinary Cooperative Research Program 
of Center for Computational Sciences, University of Tsukuba. A part of the calculation employed OpenQCD system
This work is supported in part by Grants-in-Aid for Scientific Research from the Ministry of Education, Culture, Sports, 
Science and Technology (MEXT) (Nos.16H06002, 18K03638, 19H01892).


\begin{thebibliography}{99}
\bibitem{CKM} 
M. Kobayashi and T. Maskawa, Prog. of Theoretical Physics. 49 (2) : 652-657.
\bibitem{PDG}
Particle Data Group (M. Tanabashi {\it et al.}), Phys. Rev. D 98, 030001 (2018).
\bibitem{FLAG}
FLAG working group (S. Aoki {\it et al.}), arXiv:1902.08191.
\bibitem{Vus}
PACS collaboration (J. Kakazu. {\it et al.}), arXiv:1912.13127.
\bibitem{PACS10}
PACS collaboration (N. Ukita {\it et al.}), Phys. Rev. D 99, 014504 (2019). 
\bibitem{ref.conf1}
C. Morningstar {\it et al.}, Phys.Rev.D69,054501 (2004).
\bibitem{ref.conf2}
Y. Iwasaki, arXiv:hep-lat/1111.7054v1.
\bibitem{ffL3}
RBC/UKQCD collaboration (P. A. Boyle  {\it et al.}), JHEP 1506 (2015) 164.
\bibitem{ref.wrap}
PACS collaboration (J. Kakazu. {\it et al.}), PoS LATTICE2016 (2017) 160.
\bibitem{ChPT}
J. Gasser and H. Leutwyler, Nucl. Phys. B250, 517 (1985)
\bibitem{conf96}
PACS collaboration (N. Ukita {\it et al.}), PoS LATTICE{ 2015}  (2016) 075 .
\bibitem{ZV}
PACS collaboration (K.I. Ishikawa  {\it et al}), PoS LATTICE{ 2015}  (2016) 271.
\bibitem{ffL6}
JLQCD collaboration (S. Aoki {\it et al.}), Phys. Rev. D96, 034501 (2017).  
\bibitem{ffL1}
ETM collaboration (V. Lubicz {\it et al.}), Phys.Rev.D80:111502 (2009) .
\bibitem{ffL2}
ETM collaboration (N. Carrasco {\it et al.}), Phys. Rev. D 93, 114512 (2016).
\bibitem{ffL4}
FNAL/MILC collaboration (A. Bazavov {\it et al.}), Phys.Rev. D87 (2013) 073012.
\bibitem{ffL5}
FNAL/MILC collaboration (A. Bazavov {\it et al.}), Phys.Rev. D99 (2019) no.11, 114509 .
\bibitem{ffexp}
M. Moulson, PoS CKM2016, 033 (2017).
\bibitem{Vud} 
J. C. Hardy and I. S. Towner, PoS CKM2016, 028 (2016).
\bibitem{Vud2} 
C. -Y. Seng {\it et al.}, Phys. Rev. Lett. 121, 241804 (2018).
\end{thebibliography}
\end{document}